\documentclass[reprint,twocolumn,amsmath,amssymb, aps,pra,superscriptaddress]{revtex4-1}

\usepackage{graphicx}
\usepackage{dcolumn}
\usepackage{physics}
\usepackage{bm}
\usepackage{bbold}
\usepackage[usenames,dvipsnames]{color} 
\usepackage[colorlinks=true , citecolor=blue,urlcolor=blue]{hyperref}

\newcommand{\ftN}{$^{14}$N }

\begin{document}
\title{Cross-sensor feedback stabilization of an emulated quantum spin gyroscope
}

\author{J.-C. Jaskula}\thanks{these authors contributed equally to this work}
\affiliation{Department of Nuclear Science and Engineering and Research Laboratory of Electronics, Massachusetts Institute of Technology, Cambridge, MA, USA}
\author{K. Saha}\thanks{these authors contributed equally to this work}
\affiliation{Department of Nuclear Science and Engineering and Research Laboratory of Electronics, Massachusetts Institute of Technology, Cambridge, MA, USA}
\affiliation{Department of Electrical Engineering, Indian Institute of Technology Bombay, Mumbai 400 076, India}
\author{A. Ajoy}
\affiliation{Department of Nuclear Science and Engineering and Research Laboratory of Electronics, Massachusetts Institute of Technology, Cambridge, MA, USA}
\affiliation{Department of Chemistry, University of California Berkeley, and Materials Science Division Lawrence Berkeley National Laboratory, Berkeley CA}
\author{D. J. Twitchen}
\affiliation{Element Six Innovation, Fermi Avenue, Harwell Oxford, Didcot, Oxfordshire OX11 0QR, United Kingdom}
\author{M. Markham}
\affiliation{Element Six Innovation, Fermi Avenue, Harwell Oxford, Didcot, Oxfordshire OX11 0QR, United Kingdom}
\author{P. Cappellaro}\email{pcappell@mit.edu}
\affiliation{Department of Nuclear Science and Engineering and Research Laboratory of Electronics, Massachusetts Institute of Technology, Cambridge, MA, USA}

\begin{abstract}
Quantum sensors, such as the Nitrogen Vacancy (NV) color center in diamond, are  known for their exquisite sensitivity, but their performance over time are subject to degradation by environmental noise.
To improve the long-term robustness of a quantum sensor, 
here we realize an integrated combinatorial  spin sensor in  the same micrometer-scale footprint, which exploits two different spin sensitivities to  distinct physical quantities 
to stabilize one spin sensor with local information collected in realtime via the second sensor. 
We show that we can use the  electronic spins of a large ensemble of NV centers  as  sensors of the local magnetic field fluctuations, affecting both spin sensors, in order to stabilize the output signal of interleaved Ramsey sequences performed on the \ftN nuclear spin. An envisioned application of such a device is to sense rotation rates with a stability of several days, allowing navigation with limited or no requirement of geo-localization. Our results would enable stable rotation sensing for over several hours, which already reflects better performance than MEMS gyroscopes of comparable sensitivity and size.  
\end{abstract}

\maketitle

\section{Introduction}
Our quest to understand the fundamental laws of nature and to design ever more advanced technologies requires precise measurements with outstanding performance even in challenging environmental conditions. 
Realizing  breakthrough discoveries and revolutionary technologies, such as gravitational wave detection and self-driving cars, 
often implies measuring  extremely weak signals, demanding a continuous improvement  of our measurement tools. Two figures of merit, sensitivity and stability, are crucial for these tasks: while quantum sensors have achieved sensitivities beyond any other technology, they are often prone to instability and decoherence due to external influences. One such quantum sensor  is the Nitrogen Vacancy (NV) center in diamond, which is one of the most promising platforms for quantum sensing and many other applications of quantum mechanics~\cite{RN2, RN14}. 
Increasing the coherence of NV centers via better controlled growth of diamonds~\cite{Balasubramanian2009}, implementation of dynamical decoupling sequences~\cite{Bar-Gill2013,deLange2010,Knowles2014}, and quantum memories~\cite{RN21} are few of the many advances that have led to an improved  magnetic field sensitivity,  able to probe nanoscale weak phenomena in condensed matter~\cite{RN12, RN17, RN18} and biology~\cite{RN19}. Measuring weak signals is however not just a matter of using a sensitive device, but also being able to extract signals out of environmental noise via long averaging. In turns, this requires  using stable sensors as well as implementing protocols to suppress the effects of different noise sources.

As NV centers comprise an electronic  and a nuclear spin within a single lattice site~\cite{RN3}, they enable a broader range of potential applications. Here we report the  design of a compact combinatorial device containing two different large ensembles of sensors in the same footprint, taking advantage of the  very high densities ($\sim\!10^{17}$~cm$^{-3}$) of solid-state systems.  We demonstrate a cross-sensing application of the NV electronic and nuclear spin, where the nuclear spin is used as the primary sensor, while the electronic spin, by sensing the exact same fluctuations of the environment, is used to stabilize it. Specifically, we implement Ramsey interferometry with a nuclear spin ensemble, a protocol  that could be exploited to make a rotation sensor. The operating principle of such a quantum spin gyroscope is based on the detection of the dynamic phase accumulated due to the rotation of a spin around its symmetry axis~\cite{RN4}. Alternative NV spin gyroscope designs are based on the measurement of the Berry phase~\cite{RN5, RN22, RN26}, the shift of the Larmor frequency of $^{13}$C nuclear spins due to pseudo-fields~\cite{Wood17}, or an effective AC magnetic field when the spins are rotating in a non-coaxial static magnetic field~\cite{RN6}. The two latter techniques rely on the ability of the NV centers to measure their magnetic field environment, which requires to use the highly sensitive NV electronic spins that suffer from shorter coherence time compared to the \ftN nuclear spins. On the other hand, gyroscopes probing the geometric phase due to the adiabatic evolution of the Hamiltonian during a rotation will have similar performance as devices that measure the dynamic phase. Both can take advantage of using a large number of \ftN nuclear spins with long dephasing time in an isotopically purified $^{12}$C diamond to promise rotation sensitivities of the order of 10$^{-1}$~$\mathrm{\ deg\  s^{-1}/\sqrt{Hz}}$~\cite{RN5, RN22, RN26}. 
Indeed, because of its small gyromagnetic ratio ($\gamma_N\!=\!0.3$~kHz/G), the \ftN nuclear spin is a poor magnetic field sensor, so it is less perturbed by any magnetic environmental noise and hence exhibits a coherence time of $\sim\!1$ ms in large ensembles. 

To improve on that, we will pair our rotation sensing protocol with a feedback loop to reach long-term stability. Typically, feedback protocols as for instance quantum error correction (QEC) codes are implemented via ancilla qubits either for their isolation to the environment or to use redundant degrees of freedom. QEC codes have been proposed with NV centers for quantum metrology \cite{Kessler14,Unden16} where the NV electronic spin is used as main sensor and nearby nuclear spins as ancilla qubits. Here our approach is the opposite: we rely on the advantages of the nuclear spins as presented above and exploit the fact that the NV spin is instead very sensitive to its environment. 
We quantify the stability of our quantum sensor in the presence of a controlled magnetic perturbation, in both a free running and a corrected regime, by computing the Allan deviation, which highlights characteristic features related to the different types of noise affecting our sensors. In particular, we demonstrate using this figure of merit that we can stabilize the output signal of the nuclear spins via an active feedback scheme using the NV electronic spin as a local magnetic field sensor to monitor the common environment of both spins. We thus recover a square root behavior of the Allan deviation, enabling an efficient averaging of the nuclear Ramsey signal over a period longer than a day.

\section{Results}
Our device is based on an ensemble of NV$^-$ defects in diamond (see Appendix~\ref{sec:fptESR}), providing a hybrid electron-nuclear spin system with optical addressability. 
We designed optical and microwave control apparatus, as well  as control protocols, in order to demonstrate the capabilities of this sensor. 
By applying a green laser beam focused on a 10 $\mu$m spot during 30 $\mu$s, we initialize $10^9$ NV centers in the $m_S=0$ electronic spin ground state manifold.  
Two permanent magnets in a Helmholtz configuration apply a static field of 420~G,  aligned along the NV-center's $\langle111\rangle$ axis.
Coherent control between the $m_S=0$ and $m_S = -1$ spin states is performed via pulsed resonant microwave at 1.7 GHz. The longitudinal component of the NV spin state can be determined optically  by monitoring the fluorescence intensity with 3~PIN photodiodes placed in contact with the edges of the diamond and thus collecting 6\% of the total fluorescence ~\cite{RN8}. 

\begin{figure*}[tbh!]\centering
\includegraphics[width=0.8\textwidth]{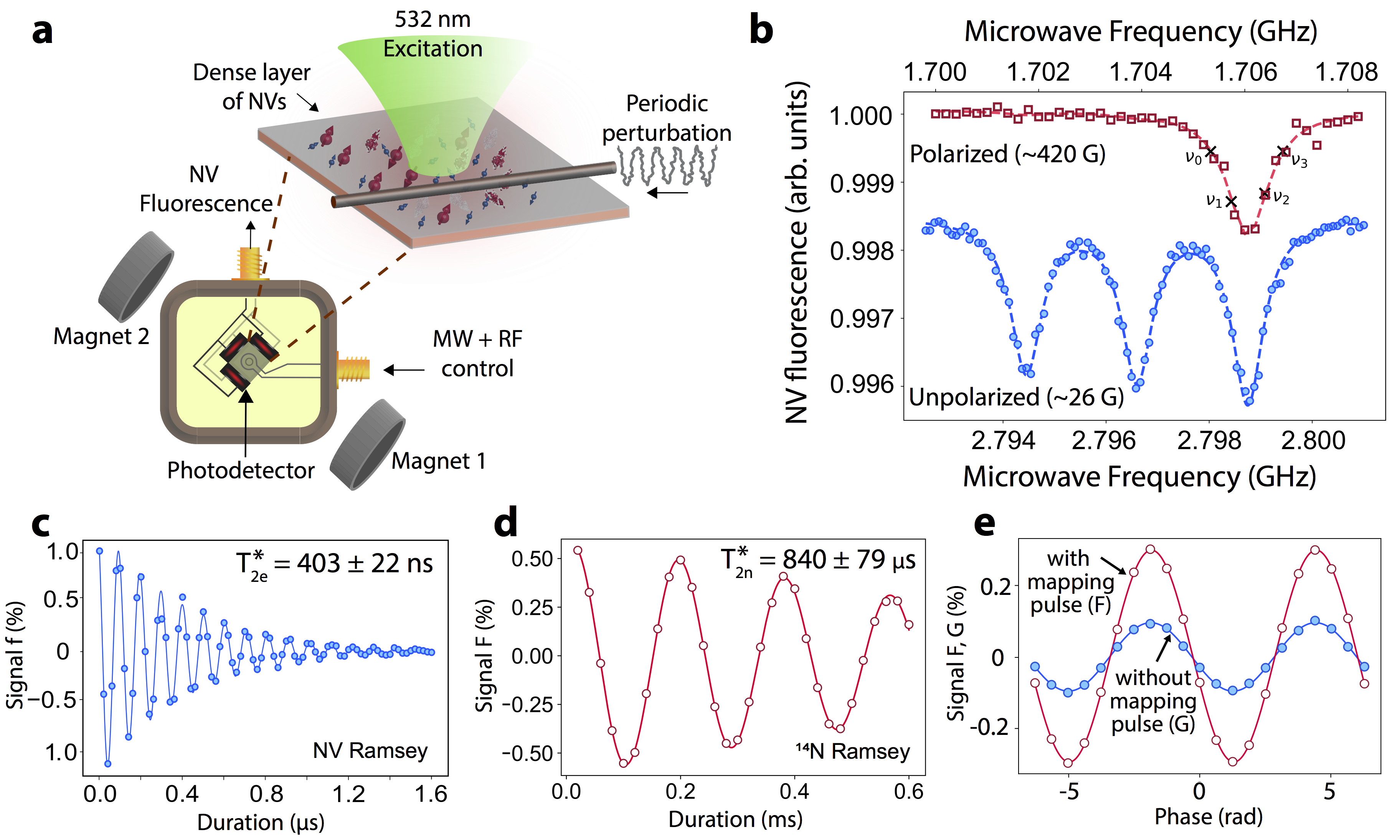}
\caption{\textbf{Control and readout of spin ensembles. }
(a) Schematics of the experimental device (See main text for a description). (b) Pulsed electronic spin resonance (ESR) spectrum of an ensemble of NV centers. The fluorescence signal is normalized to this measured prior to driving the NV spin with microwave. We added an offset to the blue curve for better visualization. In blue, NV centers are aligned with an external magnetic field of $\sim 26$~G and show an equal population in each nuclear spin state. On the other hand, for a magnetic field of $\sim 420$~G (red), the $^{14}$N nuclear spin is initialized in a particular spin state ($\ket{m_S=0, m_I=1}$) via a transfer of polarization that occurs close the excited state level anti-crossing (ESLAC). 
(c) Coherence decay of the electronic spin. We use two subsequent Ramsey sequences, each one composed of two $\pi/2$ pulses detuned from the resonance frequency $\nu_e$~=~1.704 GHz by $\Delta \nu_e$~=~10~MHz. The phase of the second $\pi/2$ pulses are shifted by $\pi$ to measure both spin projections. Similarly to equation \eqref{eq:sigF}, we plot the signal difference f. This signal oscillates at the detuning frequency $\Delta \nu_e$ and decays with a coherence time  T$_{2e}^*$ = 403(22)~ns.
(d) Coherence decay of the nuclear spin. The resonance frequency $\nu_n$ is 4.68  MHz and the detuning $\Delta \nu_n$~=~5.5~kHz. The nuclear spin is prepared in a superposition of state $(\ket{0,0} +\ket{0,1} )/2$. We measure a coherence time T$_{2n}^*$ = 840(79) $\mu$s via a Ramsey pulse sequence. A microwave $\pi$ pulse tuned at $\nu_e$ is used a selective mapping ($\ket{0,0} \rightarrow \ket{-1,0}$, $\ket{0,1} \rightarrow \ket{0,1}$) between the nuclear spin state and the NV electronic spin state to obtain higher contrast during the spin-dependent fluorescence readout. (e) Contrast response of the nuclear spin sensor to a linear change of the phase of the last $\pi$/2 pulse of the Ramsey sequence with (without) selective mapping in red (blue, Appendix~\ref{sec:raw}). This is the same response as if the NV sensor would be accumulated a quantum phase $\Phi$ during a time t = 600~$\mu$s due to physical rotation about the NV axis. The nuclear spin readout can be done without a mapping releasing the constraint of a narrowband, selective pulse but with lower contrast.
}
\label{fig:fig1}
\end{figure*}
Electron  spin resonances are optically detected by sweeping the carrier frequency of a 5$\mu$s-long microwave pulse. We show in Fig.~\ref{fig:fig1}(b) the spectrum recorded from the ensemble of NV centers aligned with a magnetic field of 420~G (red) and of 26~G (blue). The hyperfine coupling between the NV center and the Nitrogen nuclear spin splits each electronic manifold into three non-degenerate states. The relative amplitude of each of the three Lorentzians of the fit provides an estimate of the degree of polarization of the nuclear spin state, which is clearly unpolarized in case of the second spectrum.  
At a magnetic field of 420~G (corresponding to a Zeeman energy $\gamma_eB\approx1.18$~GHz), a level anticrossing in the excited state (ESLAC) (zero-field splitting $\approx1.5$~GHz) allows for polarization transfer from the electronic spin onto the nuclear spin~\cite{RN9, RN25}. This results in initializing both NV center spins into the $\ket{m_S=0, m_I=1}$ 
state with a polarization of $95(4)\%$ as is visible in the red spectrum (Fig.~\ref{fig:fig1}(b)). 

We then choose a pair of hyperfine states ($\ket{0,1}$ and $\ket{0,0}$) as a basis for our sensing qubit that is coherently controlled with a radio-frequency radiation at 4.68 MHz. Reading out the difference of populations between these states is done via a selective mapping between the state $m_I = 0$ onto $m_S = -1$. Experimentally we apply a microwave pulse tuned at 1.704 GHz to be resonant on the transition $\ket{0,0}\rightarrow \ket{-1,0}$ before the fluorescence measurement. The spins that were initially in the state $m_I = 0$ will therefore fluoresce with a lower intensity ~\cite{RN11}. Our sensing qubit benefits from the longer coherence time of the \ftN nuclear spin ~\cite{RN10}, which be measured by plotting the change of NV fluorescence signal F given by equation \eqref{eq:sigF} after applying a series of Ramsey pulse sequences (described in \ref{sec:gyro}). In Figure~\ref{fig:fig1}(d), we observe a decaying signal with a nuclear dephasing time of $T_{2n}^{*} = 840(79)\ \mu$s.

\subsection{Emulated nuclear spin gyroscope}
\label{sec:gyro}
Using all these steps together we can implement quantum sensing protocols to measure various physical quantities. The Ramsey sequence 
is one simple example of such protocols where we drive the nuclear spin sensor to a superposition of state, after which it evolves freely during a precession time $t$ that is usually set on the order of the coherence time $T_{2n}^{*}$. It will thus acquire a relative phase dependent on the strength of the measured quantity that is transferred into a difference of population for optical readout. 
In the case of a rotating spin at the rate $\Omega$ and in the presence of an external magnetic field $b$, the phase is given by $\Phi = (\gamma b + \Omega) t$. In other words, a physical rotation would be coupled into this dynamic phase and mapped out through a population difference. Equivalently, and more intuitively, we can consider the two rf $\pi/2$-pulses as being applied along two different axes. The first pulse is along a reference axis (x-axis by convention) while the second one is about an arbitrary axis (x$'$-axis) rotated by an angle $\theta=\Omega t$. In our experimental proof-of-principle of rotation sensing, we emulate such a phase accumulation by cycling the phase of the last pulse of the Ramsey sequence as shown in Fig.~\ref{fig:fig1}(e), followed by a spin readout that includes a mapping pulse (red). Setting the accumulation time $t = 600\ \mu$s close $T_{2n}^*$, this sequence simulates a rotation at the rate $\Omega_s = \theta/t$. From a statistical analysis of the data of Figure~\ref{fig:fig1}(e), we determine the rotation rate sensitivity as the signal amplitude equivalent to the amount of noise $\delta \Omega$ after averaging N$_\mathrm{seq}$ subsequent sequences during a total acquisition time of one second (i. e. the averaged signal for a signal-to-noise ratio of 1). Experimentally, it is measured as $\eta = \sigma_f(T) \sqrt{T}/dS_\Omega$ where $\sigma_f$ is the standard error in a set of fluorescence signal measurements, T the measurement time and $dS_\Omega$ is the slope of the fluorescence signal as a function of the rotation rate (here $\Omega = \Phi/t$ from Figure~\ref{fig:fig1}(e)). We obtained a sensitivity of $3000 \mathrm{\ deg\  s^{-1}/ \sqrt{\mathrm{Hz}}}$ for nuclear spins. A sensitivity comparison with electronic spins (sensitivity of $0.5 \times 10^6 \mathrm{\ deg\  s^{-1}/ \sqrt{\mathrm{Hz}}}$) indeed shows that the nuclear spin sensor benefits from longer coherence time. 
We believe  that our sensitivity is reduced by technical limitations that include an excess of electrical noise from the photodetectors as well as an excess of background light from the microwave circuit that reduces the contrast. We estimated that technical improvements could lead to sensitivities in the order of 10$\mathrm{\ deg\  s^{-1}/ \sqrt{\mathrm{Hz}}}$ (See Appendix \ref{sec:setup}). Further improvement can be obtained with dynamical decoupling techniques and spin-bath driving that would allow for an extension of the coherence time T$_2^*$~\cite{Chen18, Bauch18}.

Close to the ESLAC, the mechanism of nuclear spin repolarization provides a mean to read out the nuclear spin state without using a narrowband, selective microwave pulse. Indeed, the polarization process is enabled by flip-flops of the electronic and nuclear spins in the excited state. The ensuing state swapping also modifies the measured fluorescence intensity  depending on the initial nuclear state.
Such a method has the main advantage of being intrinsic to the NV center and consequently not relying on any coherent control that requires calibration and stability. It faces however the drawback of a lower contrast, here by a factor 3 (blue, Fig.~\ref{fig:fig1}(e)), as the state swapping is stochastic in nature, due to the 12ns short excited state lifetime, and not coherent as for the selective pulse.

Our device is built in a very compact design as nanoscale, combinatorial sensors are embedded in the same footprint and coherent control can be delivered by the same loop antenna. Thus, the nuclear and electronics spins allows for measurements of two independent quantities like temperature, magnetic and electric fields~\cite{RN7} as well as rotations, probed at the same lattice site. Besides, the two sensors are then sensitive to the same local environment that could cause errors and drifts of their output signal at different timescales. 
In particular, in our setup, magnetic field amplitude drifts due to a change of magnetization of the permanent magnets will affect both sensors similarly: a change of magnetic field strength will induce a phase shift during the Ramsey sequence and will cause systematics in the rotation measurement. In the following we analyze the stability of our combinatorial quantum sensor and describe schemes to mitigate drifts.

\begin{figure*}[thb]\centering
\includegraphics[width=0.85\textwidth]{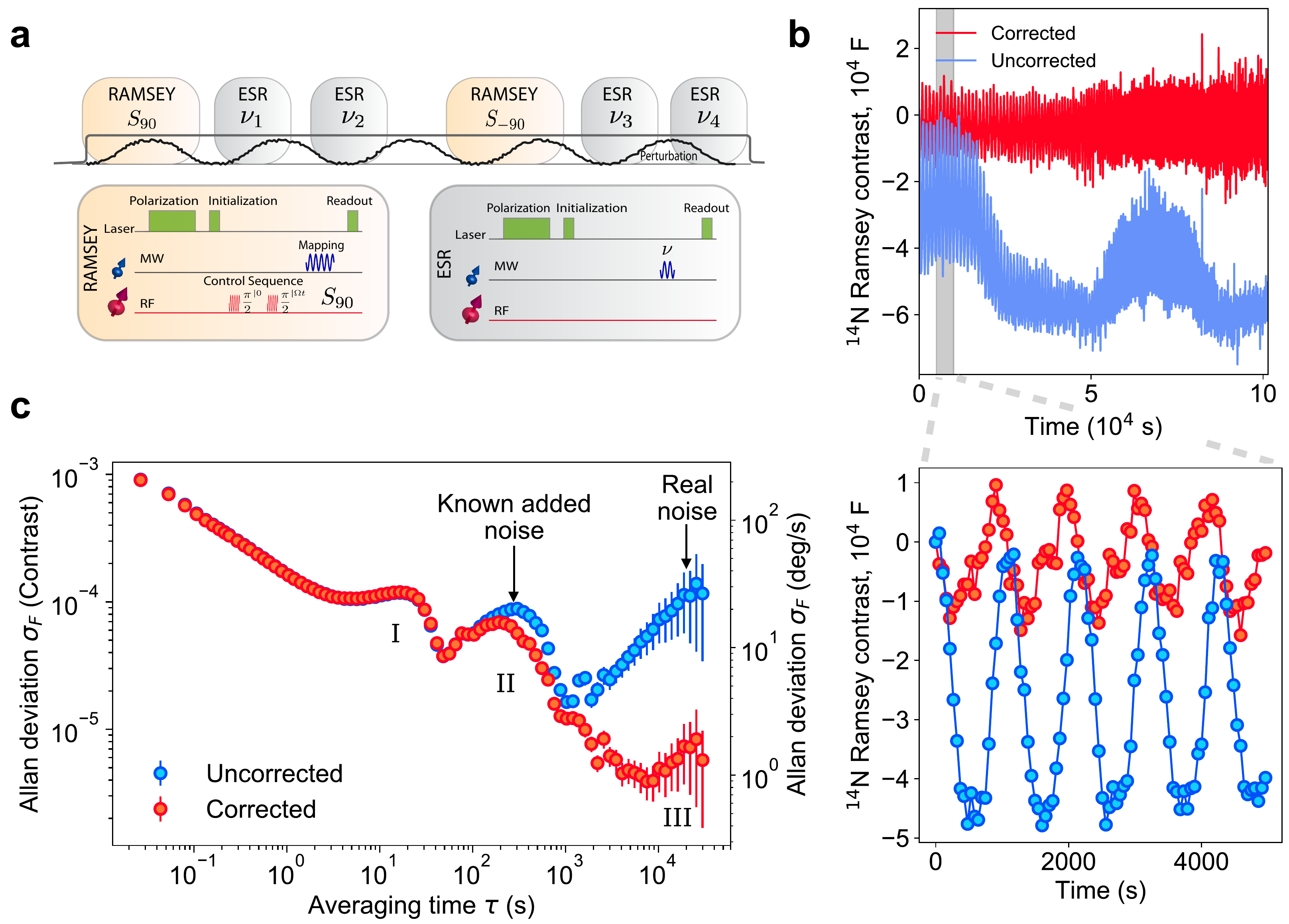}
\caption{\textbf{Nuclear spin sensor with stabilized readout.} (a) The sequence consisting in alternating measurements of the quantum phase via a Ramsey sequence with the \ftN nuclear spins ($t=600 \mu$s) and measurements of the energy shift via a set of ESR measurements with the NV spins. The full sequence is symmetrized to reject common noise. (b) Nuclear Ramsey contrast F as defined by equation \eqref{eq:sigF} with an uncorrected (blue) and corrected (red) selective mapping. A slowly varying magnetic field shifts the energy levels and perturb the mapping. (c) Allan deviations of the Ramsey signals of (b). The maximum averaging time $\tau$ used to calculate the Allan deviation is about a third of the total acquisition time of 1.15 day (see Appendix~\ref{sec:AllanApp}). The right axis is rescaled using a factor s = $4.4\times 10^6\ (\mathrm{deg\  s^{-1}})^{-1}$ calculated at the steepest point of curve F in Figure~\ref{fig:fig1}(e).}
\label{fig:fig2}
\end{figure*}
Here we design and implement an adaptive protocol ~\cite{RN12, RN13,RN13b, RN20} to locally probe magnetic field changes and feed this information back on both sensors to stabilize their combined output signal. 
The control protocol is depicted in Figure~\ref{fig:fig2}(a) and consists in six interleaved measurements. 
We apply a Ramsey sequence  to the nuclear spins. The relative phase between the two $\pi/2$ pulses is chosen to $\theta_r=90^{\circ}$ so the nuclear signal is 0 when no phase is accumulated (see Figure~\ref{fig:fig1}(e)). As described above, a phase shift of the driving field around this ideal bias point mimics a physical rotation that we want to detect.  
In a regime of small signals, the response of this first sensor will be a linear change of the fluorescence output signal $S_{90}$, measured with the spin readout which includes a mapping step as described above. 
This sensing module is followed by two spin resonance measurements at the frequencies $\nu_4 = \nu+700$~kHz and $\nu_3= \nu+350$~kHz. The second half of the sequence consists in repeating a similar set of measurements with a relative phase $\theta_r = -90 ^\circ$ and the ESR frequencies $\nu_1 = \nu-700$~kHz, $\nu_2= \nu-350$~kHz. All these frequencies are graphically represented in Figure~\ref{fig:fig1}(b). In a regime where the measured physical quantities are slowly varying with respect to the total sequence length ($\sim 1$ ms), noise that have a similar signature on both output signals $S_{\pm90}$, as for example laser intensity noise, can be suppressed by using the effective signal 
\begin{equation}
F = \frac{S_{90} - S_{-90}}{S_{90} + S_{-90}}.%
\label{eq:sigF}
\end{equation}
However, such a common-noise rejection scheme is still inefficient in case of sources of noise, such as magnetic fields, that act similarly to a signal. To suppress them, we exploited the four ESR measurements~\cite{RN1} to probe line shifts caused by these sources of noise, interleaved with the sensing protocol of the first sensor. The relative fluorescence intensity at four different microwave frequency allows for recovering the transition frequency $\nu$ and determining the strength of the field causing this line shift (see Appendix~\ref{sec:fptESR}). 

In the following, we test our stabilizing schemes against an engineered, slowly drifting perturbation generated by applying an oscillating magnetic field created by a coil placed at 1~cm of the diamond sample. Its period is set to 1000~s and its strength along the NV axis is measured to be 0.14 G peak to peak via the four-point ESR measurements described above. In Figure~\ref{fig:fig2}(b), a clear oscillation of a period of 1000~s is visible in the signal of the nuclear Ramsey measurements (blue data points) as well as a contribution from a slower environmental noise, on which we have no control, on a timescale of a few hours. While nuclear spins are little sensitive to our applied magnetic perturbation, NV spins are far more affected by it, which  highly disturbs the mapping step. Indeed, the transition frequency of selective pulses must be finely calibrated to maximize the readout fidelity and be stable along the full measurement dataset in order to limit readout errors. The feedback protocols we implemented however succeed in  stabilizing both the nuclear and electronic spin transition frequency fluctuations. In what follows, we present two scenarios in which we isolate the effect of the perturbation to a single parameter that will be stabilized. First, we compensate the readout mapping pulse frequency to prevent a loss of contrast due to an off-resonance selective pulse. Then, while applying a stronger perturbation but no mapping, we use the signal of the electronic spin to adjust the nuclear spin driving frequency.

\subsection{Cross-sensor feedback stabilization}
We demonstrate here that we can use measurements of the local NV electronic spin to feedback on the nuclear Ramsey measurements to stabilize their results. To do that, we (i) repeat the sequence of Figure~\ref{fig:fig2}(a) N$_r$ = 2000 times, (ii) transfer the measurements on the control computer to compute the ESR shift and (iii) update the experimental parameters to compensate the measured magnetic drifts. The two last steps takes about 50~s, i.~e. for a total duration of 1~min, which would optimally set a lower bound for N$_r$  as one would like to maximize the duty cycle of the sensors. On the other hand, the characteristic timescale and amplitude of the noise limits the number N$_r$ of repetitions after which the correction has to be made as the frequency drift becomes significant. In the case of the engineered perturbation, we choose  N$_r$ = 2000 as it corresponds to the maximal drift equivalent to a tenth of the Rabi frequency of the mapping pulse and it is smaller than the bandwidth of the ESR measurements ($\sim$ 500 kHz). We show that feedback helps making the measurement more stable over the full data acquisition of more than 1 day (Figure~\ref{fig:fig2}(b), red).

More quantitatively, we characterize the stability of our dual spin sensor by computing the Allan deviation of the data traces of Figure~\ref{fig:fig2}(b) (See Appendix~\ref{sec:AllanApp} and Figure~\ref{fig:fig2}(c)). We observe that the uncorrected signal (shown in blue) displays an overall decaying behavior with three features. The first two at T=50~s and T=1000~s are the signature of periodic noises at frequencies 1/$T$. They correspond to perturbations associated with (I) the episodically interrupted recordings to update the experimental parameters and (II) the magnetic perturbation that we apply to our sensor. The third noticeable feature (III) is due to the environmental noise that prevents sensors to operate accurately over long runs. We believe this is due mainly temperature changes that affect the magnetization of the permanent magnets. Shown in red is the Allan deviation for the corrected data set which displays a varying stability improvement at different timescale. The perturbation (II) is only partially corrected, mainly because of the comparable timescale of the data acquisition (about 1~min) and the magnetic perturbation period (1000~s), so that the measured field has already considerably changed by the time the correction is applied during the next acquisition. 
This is not the case for the uncontrolled environment noise (III), as its variations are much slower: then the feedback protocol based on monitoring a second spin sensor 
allows for improving the first sensor readout stability by an order of magnitude. While in this experiment we limited the feedback correction to the electronic spin driving frequency, we show next that we can obtain additional gain by directly correcting the nuclear spin control.

\begin{figure}[htb]\centering
\includegraphics[width=\columnwidth]{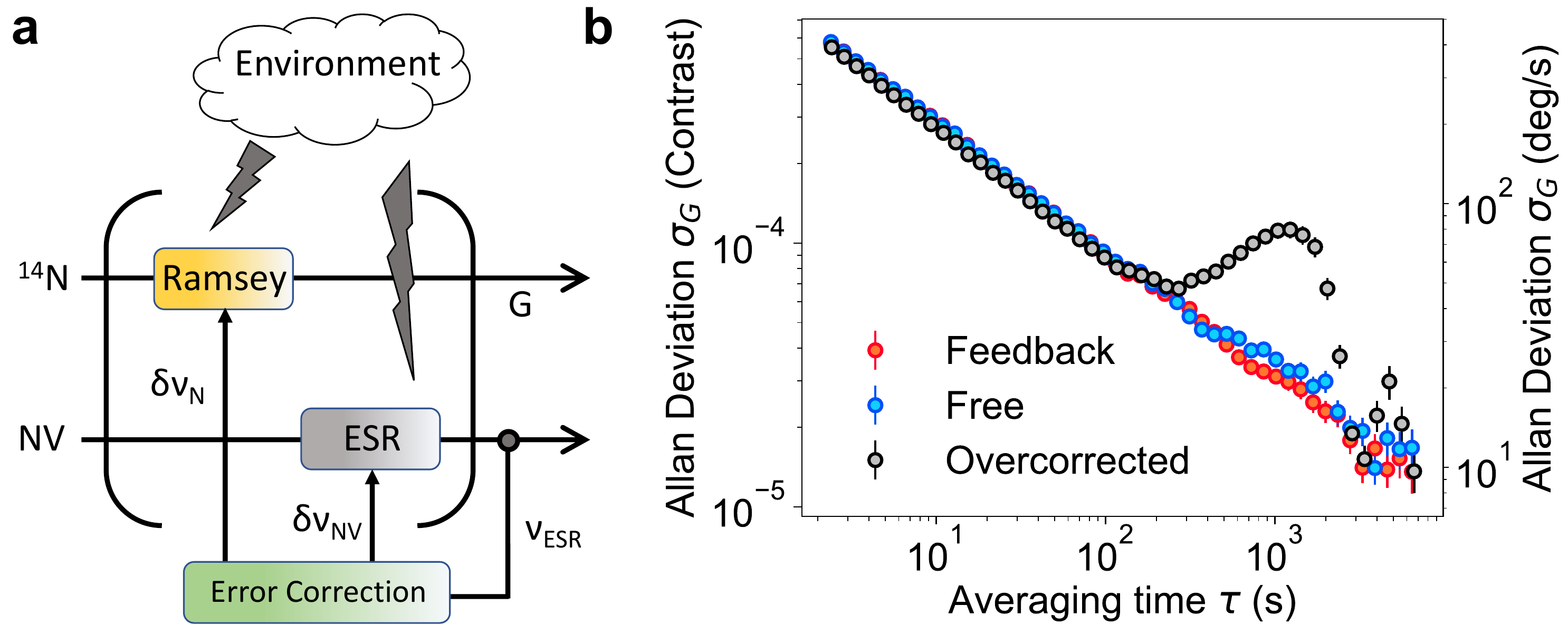}
\caption{\textbf{Stabilized nuclear spin sensor.} (a) Dual sensor scheme. The ESR shift $\nu_\mathrm{ESR}$ collected from the NV spins serves also to feedback on the nuclear spin control parameters $\delta \nu_\mathrm{N}$ to stabilize the $^{14}$N nuclear Ramsey contrast G with respect to a noise common to both spins. (b) Allan deviation of the nuclear spin signal for different correction strengths. We directly collect the nuclear spin-state dependent fluorescence without mapping pulse to isolate the perturbation effect on the nuclear spins signal. The right axis is rescaled using a factor s = $1.4\times 10^6 \ (\mathrm{deg\  s^{-1}})^{-1}$ calculated at the steepest point of curve G in Figure~\ref{fig:fig1}(e). }\label{fig:fig3}
\end{figure}
As both sensors are spins, they are sensitive to magnetic field fluctuations through Zeeman coupling. Due to a small gyromagnetic ratio (10000 times smaller than the one of an electron), the \ftN nuclear spin's response to magnetic fluctuations is weaker. Thus, to be able to see the effect of the magnetic perturbation, we increase its strength to a peak-to-peak value of $\sim$3~G. At the same time, we lengthen its period to 3000~s to stay within the limit of the previously presented four-point ESR bandwidth. Also, to isolate the effect of the perturbation on the nuclear spins, we extract the signal G directly from the bare fluorescence without any selective mapping (similarly defined as in eq. \eqref{eq:sigF}, see Appendix \ref{sec:raw}). We first plot in blue the Allan deviation of the uncorrected signal (Figure~\ref{fig:fig3}(b)). One can distinguish a small deviation from the expected square root behavior confirming that a relative phase due to the magnetic perturbation is imprinted during the free evolution of the nuclear Ramsey sequence. 

To correct this, we exploit the fact that the NV spin can probe the strength of this perturbation with a good accuracy, at exactly the same location, to cross feedback between the two sensors (Figure~\ref{fig:fig3}(a)). We probe the Zeeman shift $\delta \nu_\mathrm{NV}$ with the four-point ESR scheme and update the driving frequency of the nuclear Ramsey pulses with a frequency shift $\delta \nu_\mathrm{N} = -\gamma_N/\gamma_e \delta \nu_\mathrm{NV}$. In addition to the free running Ramsey sequence (blue), two other data acquisitions were interleaved at the same time with different corrections factors: with a the correct shift given above (red) and with its opposite (black), thus doubling the error. We can see that we recover a stable data averaging for the good feedback correction factor, whereas  the opposite correction leads to an amplification of the perturbation, thus  proving that the source of noise is indeed the same for the two spins. 
\subsection{History-based feedback protocols}
\begin{figure}[t]\centering
\includegraphics[width=0.48\textwidth]{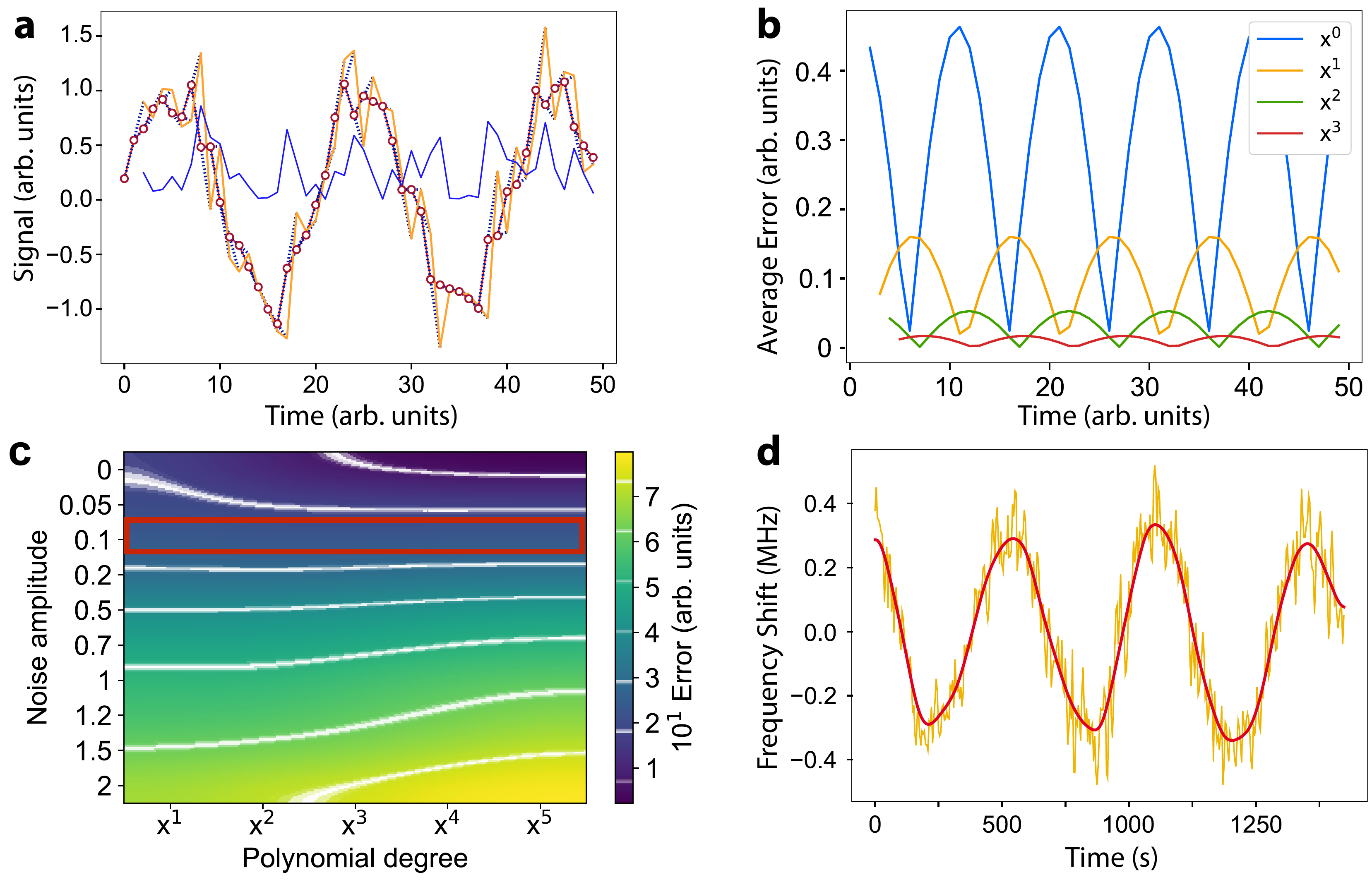}
\caption{\textbf{Simulation of history-based feedback protocols.}
(a) We simulate a sinusoidal perturbation with a stochastic noise of varying amplitude (red). A polynomial (here linear, blue dash) fit is used on certain number N of points [i,i+N] to guess the value of the point i+N+1, generating the guessed signal (orange). In blue, we plot the error between red and orange. (b) In the case of the absence of noise, we plot the error as previously defined for different degree d of the fitting polynomial. The number of points used is set N=d+1. (c) We plot the average error, defined as the mean value of the error (blue curve in (a)) averaged over 20 realizations, as a function of the noise amplitude and the degree of the polynomial. We noticed that the trend is different depending on the noise amplitude. (d) From experimental ESR data, we estimated a ratio between the magnetic perturbation and our experimental noise of 0.1. This is a regime where no improvement can be achieved using the history of the measurements as one can see in the red region of (c). }
\label{fig:fig4}
\end{figure}

So far, our feedback protocol consisted only in updating the experimental parameters with averaged values recorded during the previous dataset. However, as we keep records of every dataset, it is in principle possible to use all this knowledge to correct for slower frequency drifts with more advanced protocols. In particular, schemes relying on machine learning techniques~\cite{RN13,RN13b} or the Bayesian rule~\cite{RN12} are potential candidates to extract the most important features of the noise and be able to apply efficient corrections. Here, we would like to assess the question of the efficiency of using the previous records in the presence of stochastic noise to correct the control parameters of an ensemble of sensors.  We simulate a signal constantly equal to zero (similar to the signal F at the most sensitive operating point) on top of which is added a sinusoidal perturbation and a stochastic noise (Figure~\ref{fig:fig4}(a)). An intuitive approach to guess the best transition frequency at the $i+1$ step  is to fit the $N$ previous points with a model function, and to extrapolate the result to the future point. In the case where the stochastic noise is absent, like for example with a perfect readout, a polynomial fit allows for perfectly suppressing the perturbation, as long as one suitably increases the degree of the polynomial (Figure~\ref{fig:fig4}(b)). This is not necessary true in the scenario of a non-zero stochastic noise. As we can notice that for noise amplitude of the same order as the sinusoidal perturbation, a linear regression between the two last datasets provides a better correction than higher order polynomials (Figure~\ref{fig:fig4}(c)). Experimentally, we have measured a ratio between the magnetic perturbation and our readout noise of 0.1 (Figure~\ref{fig:fig4}(d)), indicating that we are in the regime where taking into account the past evolution of the transition frequency does not provide any help to stabilize any further our sensors. 

\section{Discussion}

We used a large ensemble of NV centers in diamond to realize a combinatorial dual spin sensor, providing the capabilities to measure two physical quantities on the same micrometer scale footprint and to stabilize one sensor with local information collected in realtime via the second sensor. Both sensors were coherently controlled in microsecond timescales with microwave and radio-frequency radiations and read out after laser excitation via an efficient fluorescence collection scheme from the side of the diamond ~\cite{RN8}. We used electron spin resonance measurements to probe the magnetic field fluctuations and stabilize the output signal of interleaved Ramsey sequences performed on the \ftN nuclear spin. Moreover, due to the strong interaction between the two spins that composed the NV center, one can use the electronic spin to increase the nuclear spin readout contrast by a factor 3. In turn, this would increase the rotation rate sensitivity by the same factor since the step of mapping extends the length of the sequence by only 5$\mu$s and doesn't affect the duty cycle. On the other hand, this mapping affects the stability of the sensor and prevent averaging beyond a certain number of repetitions. We show that our feedback scheme can improve the stability of the nuclear spin readout and the accuracy of their measurements. In figure~\ref{fig:fig2}(c), we see that the precision of the measurement tends to degrade after a total time of acquisition of $\sim 3 \tau_\mathrm{opt.}$ = 30000 s. At this optimal point, correcting the mapping pulse frequency allows to improve the precision on the averaged signal by a factor 2.5, down to a contrast error of $4 \times 10^{-6}$ (equivalent to a minimum rotation rate detectable of $\sim 1 \mathrm{\ deg\  s^{-1}}$), and almost reaches the minimum error given by a perfect average of independent measurements, which would have followed a square-root law. Given the experimental parameters, i. e. the total ESR measurement time ($2 \times 50 \mu$s) and the time to compute update the frequency (which can be reduced to less than 5 seconds), the stabilization stage does not extend significantly the sequence neither and consequently does not affect the sensitivity. Hence we believe that there is a benefit to use both mapping pulses and a stabilization procedure. 

We anticipate that such a device can potentially find application as a very stable gyroscope, allowing navigation with limited or even no need of remote localization. Existing technologies like micro electro-mechanical systems~\cite{RN101} (MEMS) or spin comagnetometers~\cite{RN23, Limes18} are already successful in making sensitive gyroscopes that have thus gained ubiquitous usage in everyday life, from navigation and inertial sensing, to rotation ion sensors in hand-held devices and automobiles. Detailed comparisons in terms of sensitivity and stability between different technologies can be find in~\cite{takase08, Passaro17, RN4}. In particular, while commercial gyroscopes achieve typical sensitivities of $0.1 \mathrm{\ deg\ s^{-1}}/\sqrt{\mathrm{Hz}}$ in hundreds of micron size footprint, their accuracy is strongly affected by drifts after few minutes of operation, making them unattractive for geodetic applications ~\cite{RN100}. On the other hand, our results show sensing capabilities for over many hours, confirming the potential of NV centers in diamond as competitive modality for such applications. Furthermore, long-term stability is a key figure of merit is the search of discrepancies in the current theories in fundamental physics and long averaging is almost always required in current tests of Lorentz and CPT symmetries or search of clues to understand dark matter~\cite{Budker14,Garcon18,Rajendran17}.

%


\appendix

\section{Experimental Methods}
All the uncertainties  represent the 95\% confidence level. 
\subsection{Experimental Setup}
\label{sec:setup}

 We use a single crystal, electronic grade (N $<$ 5 ppb) diamond substrate, with rectangular dimensions 2 mm x 2 mm x 500 $\mu$m, grown using chemical vapor deposition (CVD) by Element Six. The 13 $\mu$m thick top-surface NV sensing layer consists of 99.999\% 12C with a  \ftN concentration of 20 ppm, which has been irradiated with 4.4 MeV electrons with $1.3 \times 10^{14}$ cm$^{-2}$s$^{-1}$ flux for 5 hours and subsequently annealed in vacuum at 800 $^\circ$C for 12 hours. 
The density of NV defects is estimated~\cite{RN19} to be $2 \times 10^{17}$ cm$^{-3}$. By illuminating a 10 $\mu$m spot by  a focused green laser beam, we  address about $10^9$ spins.  The diamond is cut so that the edge faces are perpendicular to the [110] crystal axis and clamped with a PVC piece above a single 2mm-diameter, copper loop patterned on a PCB. Most fluorescence emitted by NV centers is guided via total internal reflection to the edges of the diamond chip where it is   detected by three Si PIN photodiodes (Hamamatsu S8729) that are pressed against the edges of the diamond chip. We glued  on the active area of each photodiode an high-pass optical filter ($>$ 532 nm) designed to block the leakage of excitation light and maximize the fluorescence contrast. The large 2 mm x 3.3 mm active area of the photodiodes and the short stand-off distance ($<$ 1~mm) between the sensor and the front of the optical filter ensure that the photodiode is able to collect about 6\% of optical signal within a wide solid angle. The collected fluorescence rate is measured to $5 \times 10^{13}$ photons/s, which leads to a photon-shot noise limited magnetic field sensitivity of the order of 100 pT/$\sqrt{\mathrm{Hz}}$. Increasing the beam size would allow to collect stronger fluorescence and reach magnetic sensitivities of 1~pT/$\sqrt{\mathrm{Hz}}$ or lower, comparable to those reported in~\cite{RN2, RN8, RN19}. Translated in terms of rotation rates, they would be equivalent sensitivities of 10~$\mathrm{\ deg\  s^{-1}/ \sqrt{\mathrm{Hz}}}$ with electronic spins and 0.1~$\mathrm{\ deg\  s^{-1}/ \sqrt{\mathrm{Hz}}}$ with nuclear spins because of the lower gyromagnetic ratio and longer coherence time. The electrical signal is delivered by the photodiodes is amplified by a fast  High Speed Current Amplifier (Femto, DHPCA-100), and recorded with a digital-to-analog converter (DAQ) (National Instruments NI-PCI 6251).

\subsection{Four-point ESR}
\label{sec:fptESR}

We devised a scheme to take advantage of the co-location of two sensors in our device by alternating rotation sensing (by Ramsey sequences on the nuclear spins) with a transition frequency detection via the electronic spin. The frequency depends on external environmental factors. Thus having a real-time estimate of the actual frequency allows correcting for all these parameters.

To achieve a quick estimate of the frequency we used a four-point scheme to measure the electronic spin resonance fluorescence signal at four frequencies $\nu_{1-4}$, increasingly ordered with the frequency around the expected one. To maximize the sensitivity to magnetic field changes, we set the microwave power to maximize the contrast while keeping the linewidth narrow, which results in maximizing the slope of the spectrum profile. The four frequencies are chosen as a trade-off between following the slopes at the steepest point while keeping the bandwidth large enough to track the magnetic field drifts during a complete acquisition window. The new frequency is estimated as the intersection of the lines passing by the measurements 1,2 and 3,4. Once it is determined, we use the information to correct the nuclear spin readout signal (feedback). In addition, we use this new information to select the best bias point to further measure the microwave frequency for the next time interval.

In this scheme, we assumed that it is possible, e.g., to stop the rotation during the ESR measurement, so that the frequency estimate only depends on magnetic field variations.  If this is  not practical, one could still subtract the estimate of the rotation given by the nuclear spin from the measured total phase of the electronic spin (as the rates at which phase associated with rotations are acquired is the same, while they are different for magnetic fields, such a scheme would indeed allow distinguishing between these two effects).

\subsection{Raw data }
\label{sec:raw}
\begin{figure}[htb]\centering
\includegraphics[width=\columnwidth]{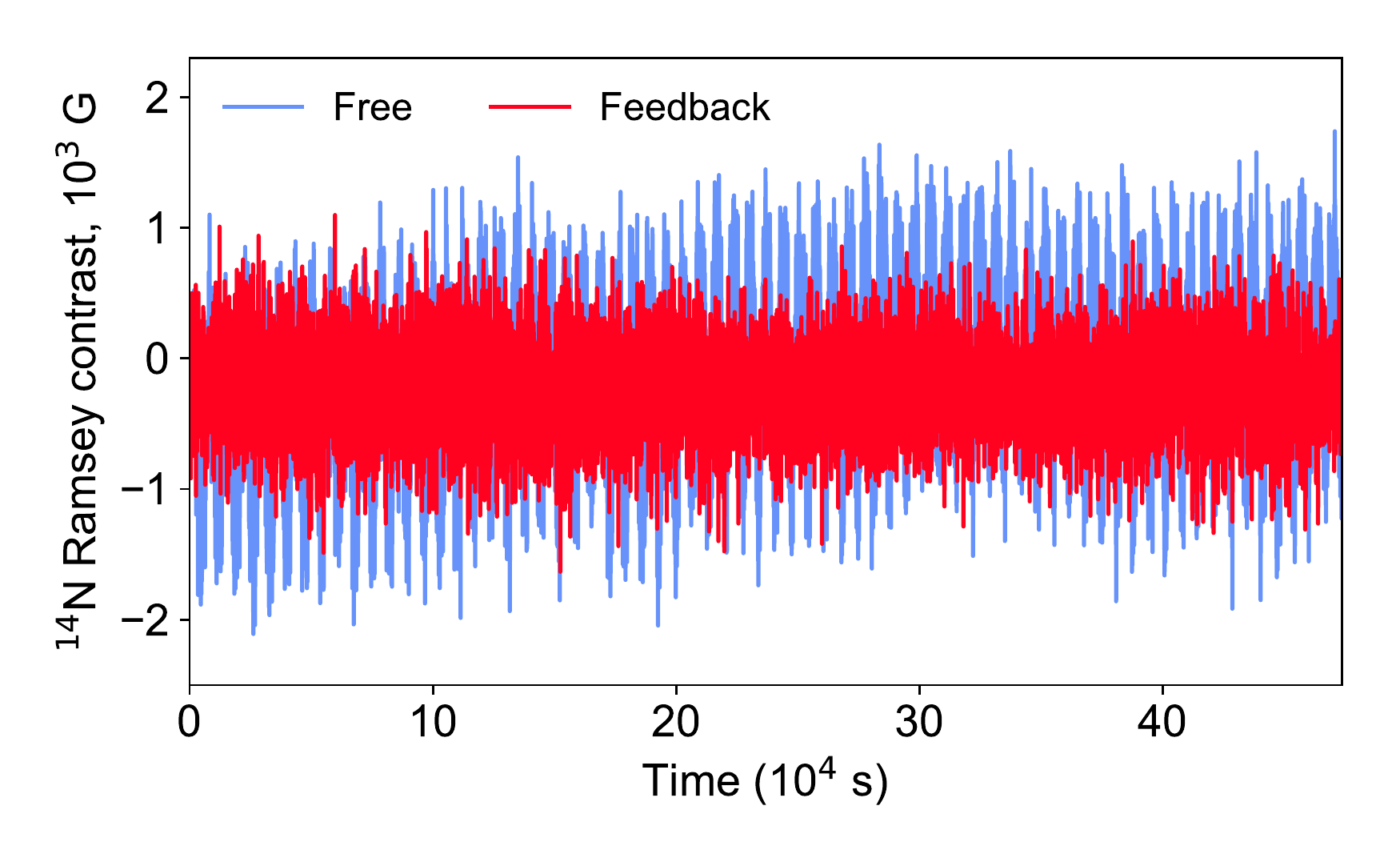}
\caption{Direct nuclear Ramsey signal.}\label{fig:figA1}
\end{figure}

We plot in figure~\ref{fig:figA1} the full data set of the nuclear Ramsey sequence. The nuclear spin readout is realized by collecting directly the fluorescence emitted by the NV center. Similarly to equation \eqref{eq:sigF}, we define the nuclear Ramsey contrast:
\begin{equation}
G = \frac{S'_{90} - S'_{-90}}{S'_{90} + S'_{-90}}.%
\label{eq:sigG}
\end{equation}
where $S'_\theta$ is the Ramsey signal recorded without using the mapping pulse (blue in figure \ref{fig:fig1}(e)). Because of flip-flop interaction at the ESLAC, the fluorescence is modulated depending on the nuclear spin state but at the third of the amplitude $S_\theta$. In figure~\ref{fig:figA1}, we see an oscillating signal at the set frequency of 0.3 mHz caused by the external arbitrary magnetic perturbation applied via a 500-turn coil. The field amplitude is measured to $\sim$3 G via an ESR made on the NV electronic spins. This strength as well as the period have been chosen such that we can notice an effect in the nuclear Ramsey while staying within the bandwidth of the 4-point ESR step. Indeed, during the acquisition of the N$_r$ = 2000 Ramsey sequences (total duration of 25 seconds including the deadtime), the line shift must stay lower than 350 kHz.

\subsection{Allan Deviation}
\label{sec:AllanApp}
The Allan variance of a signal S is defined as one half of the time average of the squares of the differences between successive readings of the frequency deviation sampled over the sampling period: 
\begin{equation}
\sigma^2(\tau) = \frac{1}{2}\left\langle \left(\frac{\Delta_\tau S}{\tau}\right)^2 \right\rangle
\end{equation}
where $\Delta_\tau S = S(t+\tau) - S(t)$. For a measurement at two intervals separated by $\tau$, the value of $\Delta_\tau S$ will be an indicator of the stability and precision of our sensor over the measured period $\tau$. Indeed, if we repeat this procedure many times, the average value of $\left(\frac{\Delta_\tau S}{\tau}\right)^2$ is equal to twice the Allan variance for observation time $\tau$ and will carry information about the noise correlations and how they affect the signal output averaging. The ideal behavior of the Allan deviation is a decay as a square root law, indicating the absence of correlation between consecutive measurements which can be then successfully averaged. At long $\tau$, it is experimentally expected that the Allan deviation starts increasing again, suggesting that the signal output is inevitably drifting due to environment noise and changes and further averaging does not improve the SNR. The error bars are directly given the number of subdivisions of the initial full dataset and their increase with $\tau$ simply reflects the fact that the number of dataset subdivision becomes inversely small for large $\tau$. Moreover, calculating an Allan deviation requires a minimum of 3 subdivisions. Thus, the maximum $\tau$ used in Figure~\ref{fig:fig2}(c) (30000~s) is at most a third of the total acquisition time of 10$^5$ seconds, i. e. 1 days and 4 hours.

\acknowledgments {This work was supported in part by the Office of Naval Research, Award No. N00014-14-1-0804, by the Army Research Office, Award No. W911NF-11-1-0400, and by  Skoltech.
}

%

\end{document}